\documentclass{pion}

\begin{document}

\markboth{} { Effect of $U_A(1)$ Breaking on Chiral Phase
Structure and Pion Superfluidity}

\catchline{}{}{}{}{}
%
%%%%%%%%%%%%%%%%%%%%%%%%%%%%%%%%%%%%%%%%%%%%%%%%%%%%%%%%%%%%%%%%%%%%

\title{Effect of  $U_A(1)$ Breaking on Chiral Phase Structure and Pion Superfluidity at Finite Isospin Chemical Potential\\ }

\author{ Lianyi He, Meng Jin and Pengfei Zhuang}

\address{Physics Department, Tsinghua University\\ Beijing 100084, China
\\
hely04@mails.tsinghua.edu.cn}

\maketitle

\begin{abstract}
We investigate the isospin chemical potential effect in the frame
of $SU(2)$ Nambu-Jona-Lasinio model. When the isospin chemical
potential is less than the vacuum pion mass, the phase structure
with two chiral phase transition lines does not happen due to
$U_A(1)$ breaking of QCD. When the isospin chemical potential is
larger than the vacuum pion mass, the ground state of the system
is a Bose-Einstein condensate of charged pions.

\end{abstract}

\ccode{PACS Nos.: .}

\section{Introduction}
It is generally believed that there exists a rich phase structure
of Quantum Chromodynamics (QCD) at finite temperature and baryon
density, for instance, the deconfinement process from hadron gas
to quark-gluon plasma, the transition from chiral symmetry
breaking phase to the symmetry restoration phase\cite{dandc}, and
the color superconductivity\cite{csc} at low temperature and high
baryon density. Recently, the study on the QCD phase structure is
extended to finite isospin density. The physical motivation to
study isospin spontaneous breaking and the corresponding pion
superfluidity is related to the investigation of compact stars,
isospin asymmetric nuclear matter and heavy ion collisions at
intermediate energies. While there is not yet precise lattice
result at finite baryon density due to the Fermion sign
problem\cite{sign}, it is in principle no problem to do lattice
simulation at finite isospin density\cite{CB1}. It is
found\cite{LA} that the critical isospin chemical potential for
pion condensation is about the pion mass in the vacuum, $\mu_I^c
\simeq m_\pi$. The QCD phase structure at finite isospin density
is also investigated in many low energy effective models, such as
chiral perturbation theory\cite{CB1,CB2,CB3}, ladder
QCD\cite{ladder}, random matrix method\cite{random}, strong
coupling lattice QCD\cite{str} and Nambu--Jona-Lasinio (NJL)
model\cite{NJL1,NJL2,NJL3,NJL3-1,NJL3-2}.

One of the models that enables us to see directly how the dynamic
mechanisms of chiral symmetry breaking and restoration operate is
the NJL model\cite{NJL4} applied to quarks\cite{NJL5}. Within this
model, one can obtain the hadronic mass spectrum and the static
properties of mesons remarkably well\cite{NJL5,NJL6}, and the
chiral phase transition line\cite{NJL5,NJL6,NJL7} in $T-\mu_B$
plane is very close to the one calculated with lattice
QCD\cite{LA2}. Recently, this model is also used to investigate
the color superconductivity at moderate baryon
density\cite{NJL8,NJL9,NJL10,NJL11,NJL12}. In this letter we
report our results on the phase structure of the general $SU(2)$
NJL model with $U_A(1)$ breaking term at finite isospin chemical
potential. At small isospin chemical potential, we argue that
there is only one chiral phase transition line in the $T-\mu_B$
plane\cite{NJL3-1}. At isospin chemical potential larger than
vacuum pion mass, the ground state of the system is a
Bose-Einstein condensate of charged pions\cite{NJL3-2}.

\section{Mean Field Approximation at Finite Isospin Chemical Potential}
We start with the flavor $SU(2)$ NJL model defined by
\begin{equation}
\label{njl1} {\cal L}
=\bar\psi(i\gamma^\mu\partial_\mu-m_0+\mu\gamma_0)\psi+{\cal
L}_{int}\ ,
\end{equation}
where $m_0$ is the current quark mass, $\mu$ the chemical
potential matrix in flavor space,
\begin{equation}
\label{mu} \mu=diag(\mu_u, \mu_d)=diag\left({\mu_B\over
3}+{\mu_I\over 2}, {\mu_B\over 3}-{\mu_I\over 2}\right)
\end{equation}
with $\mu_B$ and $\mu_I$ being the baryon and isospin chemical
potential, respectively, and the interaction part
includes\cite{NJL3} the normal four Fermion couplings
corresponding to scalar mesons $\sigma, a_0, a_+$ and $a_-$ and
pseudoscalar mesons $\eta\prime, \pi_0, \pi_+$ and $\pi_-$
excitations, and the 't-Hooft\cite{Hooft} determinant term for
$U_A(1)$ breaking,
\begin{eqnarray}
\label{njl2} {\cal L}_{int}
&=&\frac{G}{2}\sum_{a=0}^{3}\left[(\bar{\psi}\tau_a\psi)^{2}
+(\bar{\psi}i\gamma_{5}\tau_a\psi)^{2}\right]+
\frac{K}{2}\left[\det\bar{\psi}(1+\gamma_{5})\psi+\det\bar{\psi}(1-\gamma_{5})\psi\right]\nonumber\\
&=&\frac{1}{2}(G+K)\left[(\bar{\psi}\psi)^{2}+(\bar{\psi}i\gamma_{5}\vec{\tau}\psi)^{2}\right]
+\frac{1}{2}(G-K)\left[(\bar{\psi}\vec{\tau}\psi)^{2}+(\bar{\psi}i\gamma_{5}\psi)^{2}\right]\
.
\end{eqnarray}
At zero isospin chemical potential $\mu_I=0$, for $K=0$ and
$m_0=0$ the Lagrangian is invariant under $ U_B(1)\bigotimes
U_A(1) \bigotimes SU_V(2) \bigotimes SU_A(2)$ transformations, but
for $K\neq 0$, the symmetry is reduced to $U_B(1) \bigotimes
SU_V(2) \bigotimes SU_A(2)$ and the $U_A(1)$ breaking leads to
$\sigma$ and $a$ mass splitting and $\pi$ and $\eta\prime$ mass
splitting. If $G=K$, we come back to the standard NJL
model\cite{NJL5} with only $\sigma, \pi_0, \pi_+$ and $\pi_-$
mesons.

We introduce the quark condensates
\begin{equation}
\label{sigma1} \sigma_u = \langle\bar u u\rangle \ ,\ \ \ \
\sigma_d = \langle\bar d d\rangle\ ,
\end{equation}
and the pion condensate
\begin{eqnarray}
\label{pion} {\pi\over \sqrt 2} &=& \langle\bar{\psi} i\gamma_5
\tau_+\psi\rangle = \langle\bar{\psi} i\gamma_5\tau_-\psi\rangle =
{1\over \sqrt 2}\langle\bar{\psi}i\tau_1\gamma_5\psi\rangle\ ,
\end{eqnarray}
where we have chosen the pion condensate to be real. The quark
condensate and pion condensate are, respectively, the order
parameters of chiral phase transition and pion superfluidity. In
mean field approximation, the thermodynamic potential of the
system is
\begin{equation}
\label{omega} \Omega  =
G(\sigma_u^2+\sigma_d^2)+2K\sigma_u\sigma_d+{G+K\over
2}\pi^2-\frac{T}{V}\ln \det{{\cal S}^{-1}_{mf}(k)}\ .
\end{equation}
where ${\cal S}_{mf}^{-1}$ is the inverse of the mean field quark
propagator, in momentum space it reads
\begin{equation}
\label{s-1} {\cal S}_{mf}^{-1}(k)=\left(\begin{array}{cc}
\gamma^\mu k_\mu+\mu_u\gamma_0-m_u &
i\gamma_5(G+K)\pi\\
i\gamma_5(G+K)\pi& \gamma^\mu
k_\mu+\mu_d\gamma_0-m_d\end{array}\right)
\end{equation}
with the effective quark masses
\begin{eqnarray}
\label{qmass} m_u = m_0-2G\sigma_u-2K\sigma_d\ ,\ \ m_d =
m_0-2G\sigma_d-2K\sigma_u\ .
\end{eqnarray}
The condensates $\sigma_u, \sigma_d$ and $\pi$ as functions of
temperature and baryon and isospin chemical potentials are
determined by the minimum thermodynamic potential,
\begin{equation}
\label{minimum} \frac{\partial \Omega}{\partial \sigma_u}=0,\ \ \
\frac{\partial \Omega}{\partial \sigma_d}=0, \ \ \ \frac{\partial
\Omega}{\partial \pi}=0\ .
\end{equation}
It is easy to see from the chemical potential matrix and the quark
propagator matrix that for $\mu_B=0$ or $\mu_I=0$ the gap
equations for $\sigma_u$ and $\sigma_d$ are symmetric, and one has
\begin{eqnarray}
\label{sigma3}  \sigma_u = \sigma_d = \sigma/2 \ ,  m_u = m_d = m=
m_0 -(G+K)\sigma\ .
\end{eqnarray}

\section{Chiral Phase Structure at $\mu_I<m_\pi$}

We now consider the QCD phase structure below the minimum isospin
chemical potential $\mu_I^c=m_\pi$ for pion superfluidity. Since
the pion condensate is zero, there is only chiral phase structure
in this region.

It is well known that in absence of isospin chemical potential the
chiral condensate jumps down suddenly at a critical baryon
chemical potential, which indicates a first order phase
transition. At both finite baryon and isospin chemical potential,
in principle we have $\sigma_u\neq\sigma_d$, they may jump down at
the same critical baryon chemical potential or at two different
critical points. In the case $K=0$, the two gap equations for
$\sigma_u$ and $\sigma_d$ decouple and the two critical points do
exist, and therefore, there are two chiral phase transition lines
in $T-\mu_B$ plane at fixed isospin chemical potential, the
interval between the two points is just $\Delta\mu_B=3\mu_I$. What
is the effect of the $U_A(1)$ breaking term on the QCD phase
structure? If $K\neq0$, the two gap equations couple to each other
and tend to a single phase transition. Especially in chiral limit,
one can clearly see that when one of the quark condensates becomes
zero, the other one is forced to be zero for any coupling
constants $G$ and $K\ne 0$. Therefore, there is only one chiral
phase transition line at any $K\ne 0$ in chiral limit. In real
world with small current quark mass, numerical solution of the gap
equations shows that if there exists a structure of two chiral
phase transition lines depends the quantity
\begin{equation}
\label{alpha1} \alpha ={K\over G+K}= {1\over 2}\left(1-{G-K\over
G+K}\right).\
\end{equation}

In real world there are four parameters in the NJL model, the
current quark mass $m_0$, the three-momentum cutoff $\Lambda$, and
the two coupling constants $G$ and $K$. Among them $m_0, \Lambda$
and the combination $G+K$ can be determined by fitting the chiral
condensate $\sigma$, the pion mass $m_\pi$ and the pion decay
constant $f_\pi$ in the vacuum. For $\sigma = 2(-241.5$ MeV$)^3$,
$m_\pi = 140.2$ MeV, and $f_\pi = 92.6$ MeV, one has $m_0 = 6$
MeV, $\Lambda = 590$ MeV, and $(G+K)\Lambda^2/2 = 2.435$. To
determine the two coupling constants separately or the ratio
$\alpha$, one needs to know the $\eta\prime$- or $a$-meson
properties in the vacuum. In the self-consistent mean field
approximation, the $\eta^\prime$ mass $m_{\eta\prime}$ can be
calculated in random phase approximation(RPA)\cite{NJL6}
\begin{equation}
\label{pole1} 1-(G- K)\Pi_{PS}\left(k_0=m_{\eta^\prime},\bf
k=0\right)=0
\end{equation}
with polarization function $\Pi_{PS}$ defined as
\begin{equation}
\label{polari1} \Pi_{PS}(k_0,{\bf k}) = -i\int{d^4p\over (2\pi)^4}
{\bf Tr}\Big[\gamma_5{\cal S}_{mf}(p+k)\gamma_5{\cal
S}_{mf}(p)\Big].\
\end{equation}
The quantity $\alpha$ can be written as
\begin{equation}
\label{alpha2} \alpha = {1\over 2}\left(1-{1\over G+K}{1\over 12
\int{d^3{\bf k}\over (2\pi)^3}{E_k\over
E_k^2-m_{\eta\prime}^2/4}}\right)\
\end{equation}
where $E_k=\sqrt{{\bf k}^2+m^2}$. Combining with the above known
parameters $m_0, \Lambda$ and $G+K$ obtained and choosing
$m_{\eta\prime} = 958$ MeV, we have $\alpha =0.29$ which is much
larger than the critical value $0.11$ for the two-line structure
at $\mu_I = 60$ MeV\cite{NJL3}. In fact, for a wide mass region
$540$ MeV $<m_{\eta\prime}< 1190$ MeV, we have $\alpha > 0.11$,
and there is no two-line structure.

To fully answer the question if the two-line structure exists
before the pion condensation happens, we consider the limit $\mu_I
= m_\pi$. In real world with $m_0 \neq 0$ and $K\neq 0$, with
increasing coupling constant $K$ or the ratio $\alpha$ the two
lines approach to each other and finally coincide at about $\alpha
= 0.21$ which is still less than the value $0.29$ calculated by
fitting $m_{\eta\prime} = 958$ MeV. Therefore, the two-line
structure disappears if we choose $m_{\eta\prime} = 958$ MeV. In
fact, for $720$ MeV $< m_{\eta\prime} <  1140$ MeV, we have
$\alpha > 0.21$ and the two phase transition lines are cancelled
in this wide mass region. When the $\eta\prime$-meson mass is
outside this region, the $U_A(1)$ breaking term is not strong
enough to cancel the two-line structure, but the two lines are
already very close to each other. Considering the relation between
the $\eta^\prime$ mass and the number of flavors,
$m_{\eta^\prime}^2\propto N_f$, one can estimate the $\eta^\prime$
mass in the case of two flavors $m_{\eta^\prime}\approx780MeV$
which is in the region where $\alpha
> 0.21$.

\section{Pion Superfluidity at $\mu_I>m_\pi$}
\begin{figure}
\centering \includegraphics[width=2in]{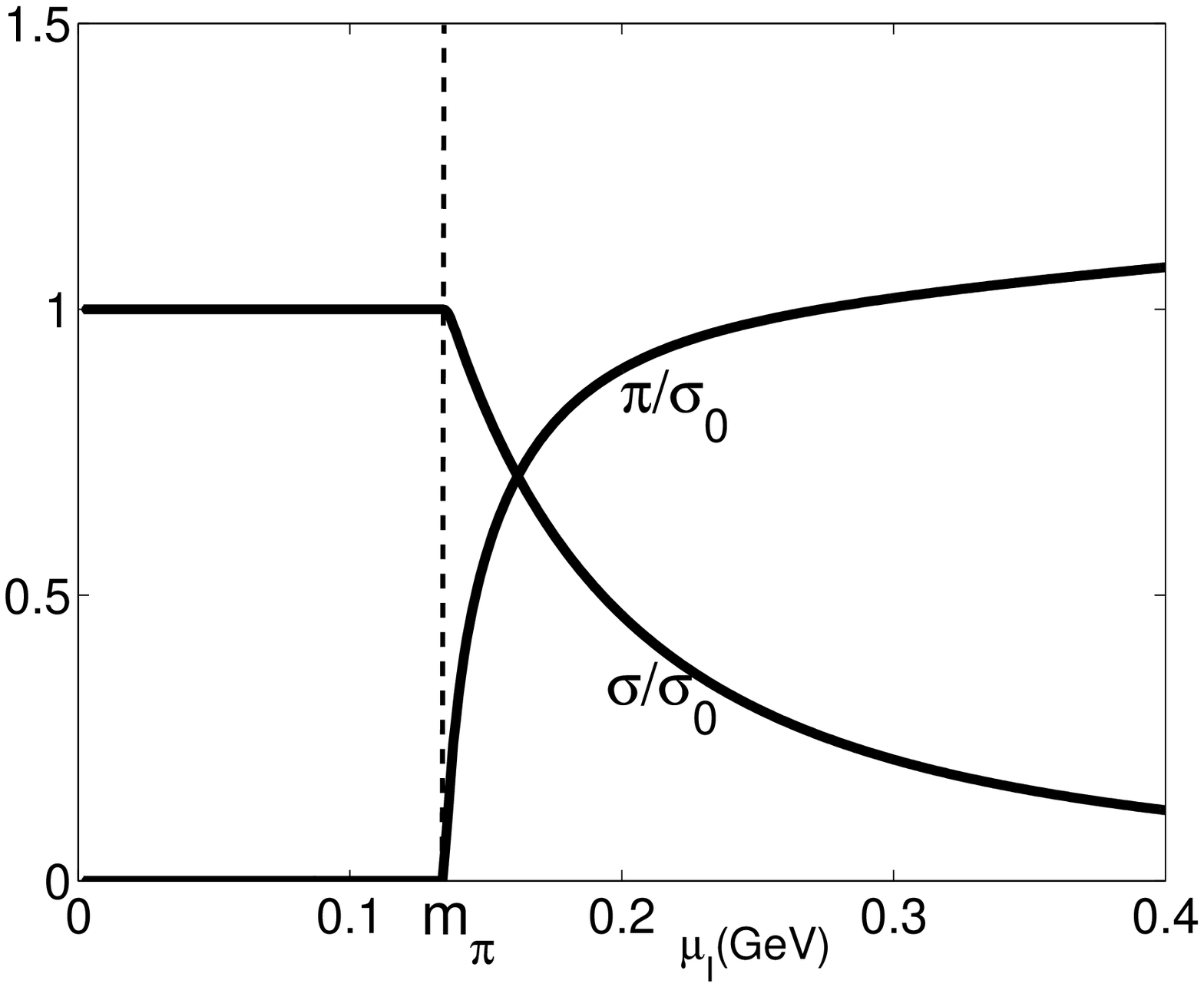}%
\hspace{0.5in}%
\includegraphics[width=2.15in]{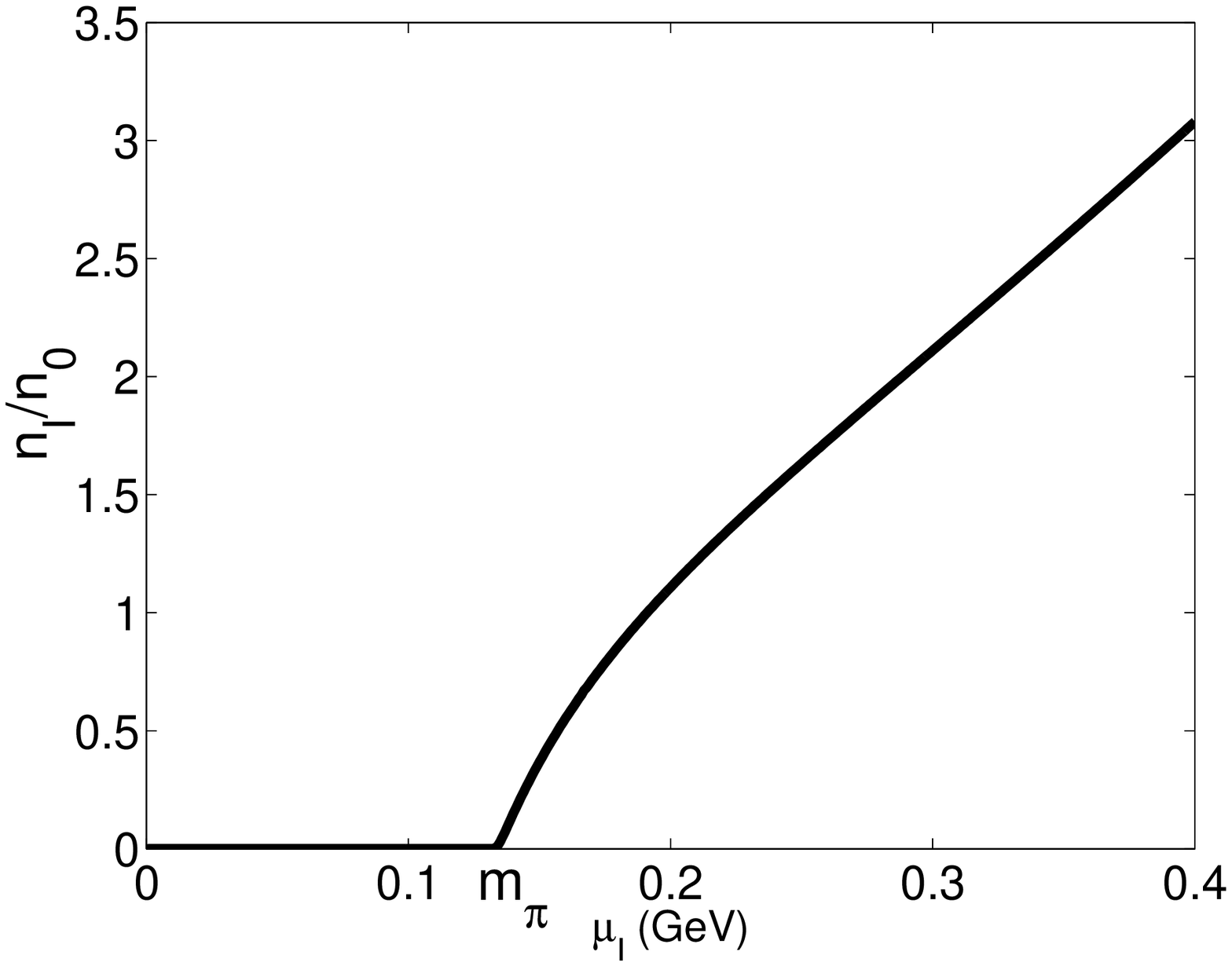}
\caption{The order parameters $\sigma$ and $\pi$, scaled by the
vacuum value $\sigma_0$, and isospin density, scaled by the value
of baryon density in normal nuclear matter.}\label{fig1}
\end{figure}

In this section we concentrate on the case $\mu_B=0$. The gap
equations for chiral condensate $\sigma=\sigma_u+\sigma_d$ and
pion condensate $\pi$ are derived as following
\begin{eqnarray}
&&\sigma = 6m \int{d^3\bf k\over (2\pi)^3}{1\over E_k}
\Bigg({E_k^-\over E_\pi^-}\left(2f(E_\pi^-)-1\right)+{E_k^+\over
E_\pi^+}\left(2f(E_\pi^+)-1\right)\Bigg)\ ,\nonumber\\
&&\pi \Bigg[1 + 12H\pi\int{d^3{\bf k}\over (2\pi)^3}\Bigg({1\over
E_\pi^-}\left(2f(E_\pi^-)-1\right)+{1\over
E_\pi^+}\left(2f(E_\pi^+)-1\right)\Bigg)\Bigg]=0\ .
\end{eqnarray}
where
$H=(G+K)/2,E_k^\pm=E_k\pm\mu_I/2,E_\pi^\pm=\sqrt{(E_k^\pm)^2+4H^2\pi^2}$.
The isospin density $n_I$ can be derived from thermodynamic
potential $\Omega$,
\begin{eqnarray}
n_I = 3\int{d^3{\bf k}\over(2\pi)^3}\left({E_k^+\over
E_\pi^+}\left(2f(E_\pi^+)-1\right)-{E_k^-\over
E_\pi^-}\left(2f(E_\pi^-)-1\right)\right).\
\end{eqnarray}
Numerical solution of the gap equations at zero temperature is
shown in Fig. \ref{fig1}. We found that when $\mu_I<m_\pi$, the
ground state is the same as the vacuum and the isospin density is
zero, while pion condensate becomes nonzero when $\mu_I>m_\pi$ and
isospin density becomes nonzero. The phase transition is of second
order. The critical isospin chemical potential $\mu_I^c$ is
exactly the vacuum pion mass $m_\pi$ which can be proved
analytically. The critical isospin chemical potential is
determined by
\begin{equation}
\label{critical} 1-12H\int{d^3{\bf k}\over (2\pi)^3}{1\over
E_k}{E_k^2\over E_k^2-\left(\mu_I^c\right)^2/ 4} = 0\ .
\end{equation}
which is just the same as the mass equation for pion in vacuum.

\begin{figure}[th]
\centerline{\psfig{file=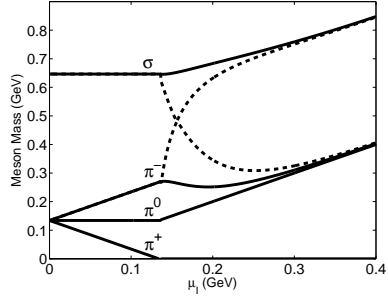,width=2.0in}} \vspace*{8pt}
\caption{Meson masses as a function of isospin chemical potential.
The solid line is full calculation and the dashed line is obtained
by neglecting $\sigma-\pi_\pm$ mixing. }\label{fig2}
\end{figure}

The collective meson excitations can be calculated in the
framework of random phase approximation. Different from normal RPA
approach in vacuum and at finite temperature, the meson modes in
the pion superfluid will mix with each other and the dispersion
relations of each eigen modes are determined by
\begin{eqnarray}
\det(1-2H\Pi(k_0,{\bf k}))=0
\end{eqnarray}
where the polarization function matrix $\Pi(k_0,{\bf k})$ is
defined as
\begin{equation}
1-2H\Pi(k_0,{\bf k})= \left(\begin{array}{cccc}
1-2H\Pi_{\sigma\sigma}&-2H\Pi_{\sigma\pi_+}
&-2H\Pi_{\sigma\pi_-}&-2H\Pi_{\sigma\pi_0}\\
-2H\Pi_{\pi_+\sigma}&1-2H\Pi_{\pi_+\pi_+}&-2H\Pi_{\pi_+\pi_-}&-2H\Pi_{\pi_+\pi_0}\\
-2H\Pi_{\pi_-\sigma}&-2H\Pi_{\pi_-\pi_+}&1-2H\Pi_{\pi_-\pi_-}&-2H\Pi_{\pi_-\pi_0}\\
-2H\Pi_{\pi_0\sigma}&-2H\Pi_{\pi_0\pi_+}&-2H\Pi_{\pi_0\pi_-}&1-2H\Pi_{\pi_0\pi_0}\\
\end{array}\right)\
.
\end{equation}
The polarization functions are defined as
\begin{equation}
\Pi_{MM^\prime}(k_0,{\bf k}) = i\int{d^4p\over (2\pi)^4}{\bf
Tr}\left[\Gamma_M^* {\cal S}_{mf}(p+k)\Gamma_{M^\prime} {\cal
S}_{mf}(p)\right]\ ,
\end{equation}
with the vertexes $\Gamma_M$ and $\Gamma_M^*$
\begin{eqnarray}
\Gamma_M = \left\{\begin{array}{ll}
1 & M=\sigma\\
i\tau_+\gamma_5 & M=\pi_+ \\
i\tau_-\gamma_5 & M=\pi_- \\
i\tau_3\gamma_5 & M=\pi_0
\end{array}\right.\ ,\ \ \ \ \ \ \
\Gamma_M^* = \left\{\begin{array}{ll}
1 & M=\sigma\\
i\tau_-\gamma_5 & M=\pi_+ \\
i\tau_+\gamma_5 & M=\pi_- \\
i\tau_3\gamma_5 & M=\pi_0 \\
\end{array}\right.\ ,
\end{eqnarray}
where $\tau_\pm=(\tau_1\pm i\tau_2)/\sqrt{2}$. From detailed
calculation one has
\begin{eqnarray}
\Pi_{\sigma\pi_\pm}=\Pi_{\pi_\pm\sigma}\propto \sigma\pi, \
\Pi_{\pi_+\pi_-}=\Pi_{\pi_-\pi+}\propto \pi^2, \ \Pi_{\pi_0
I}=\Pi_{I\pi_0}=0 (I=\sigma,\pi_\pm).
\end{eqnarray}
Thus $\pi_0$ does not mix with other modes, while
$\sigma,\pi_+,\pi_-$ mix with each other. However, mixing between
$\sigma$ and $\pi_\pm$ is only strong near the phase transition
point and can be neglect at large $\mu_I$. The meson mass spectrum
as a function of $\mu_I$ is shown in Fig. \ref{fig2}. We found
that the full calculation (solid line) is consistent with chiral
perturbation theory, and the mixing between $\sigma$ and $\pi_\pm$
plays a very important role. On the other hand, in the superfluid
phase with $\pi\neq0$, $\det(1-2H\Pi(0,{\bf 0}))=0$ is always
satisfied, which indicates that Goldstone's theorem is satisfied
in the whole superfluid phase at RPA level.
%%%%%%%%%%%%%%%%%%%%%%%%%%%%%%%%%%%%%%%%%%%%%%%%%%%%%%%%%%%%%%%%%%%%%%%
%%%%%%%%%%%%%%%%%%%%%%%%%%%%%%%%%%%%%%%%%%%%%%%%%%%%%%%%%%%%%%%%%%%%%%%

%%%%%%%%%%%%%%%%%%%%%%%%%%%%%%%%%%%%%%%%%%%%%%%%%%%%%%%%%%%%%%%%%%%%%%%
%%%%%%%%%%%%%%%%%%%%%%%%%%%%%%%%%%%%%%%%%%%%%%%%%%%%%%%%%%%%%%%%%%%%%%%%

\section{Summary}

We have investigated the two flavor NJL model with $U_A(1)$
breaking term at finite isospin chemical potential. The main conclusions are:\\
{\bf 1)} The two chiral phase transition lines in the $T-\mu_B$
plane predicated by the NJL model without $U_A(1)$ breaking term
are cancelled by the strong $U_A(1)$ breaking term at low isospin
chemical potential. Therefore, in relativistic heavy ion
collisions where the typical $\mu_I$ value is much less than
$m_\pi$, it looks impossible to realize the
two-line structure. \\
{\bf 2)} The critical isospin chemical potential for pion
condensation in NJL model is exactly the pion mass in the vacuum,
$\mu_I^c = m_\pi$, independent of the model parameters, the
regularization scheme, the $U_A(1)$ breaking term. When $\mu_I$
exceeds $m_\pi$, The ground state of the system is in a
Bose-Einstein condensate of charged pions. In this pion
superfluid, the meson modes $\sigma,\pi_+,\pi_-$ mixed with each
other and there is a Goldstone mode due to the spontaneous
breaking of $U_I(1)$ symmetry.

\section*{Acknowledgments}
The work was supported in part by the grants NSFC10135030,
10425810, 10435080,10447122 and SRFDP20040003103.


\begin{thebibliography}{0}
 \bibitem{dandc} {For instance, see Quark-Gluon Plasma,
                  ed. R.C.Hwa (world Scientific, Singapore, 1990).}
 \bibitem{csc}   {M.Alford, K.Rajagopal, and F.Wilczek, Phys.Lett.B422, (1998)247;
                  R.Rapp, T.Sch\"afer, E.V.Shuryak, and M.Velkovsky,
                  Phys.Rev.Lett.81, (1998)53.}
 \bibitem{sign}  {F.karsch, Lect.Notes Phys.583, (2002)209; Z.Fodor and S.D.Katz, Phys.Lett.B534 (2002)87; JHEP 0203 (2002)014, JHEP 0404 (2004) 050.}
 \bibitem{CB1}      {D.T.Son and M.A.Stephanov, Phys.Rev.Lett.86, (2001)592
                    ;Phys.Atom.Nucl.64, (2001)834.}
 \bibitem{LA}      {J.B.Kogut,D.K.Sinclair, Phys.Rev.D66, (2002)034505.}
 \bibitem{CB2}     {J.B.Kogut and D.Toublan, Phys.Rev.D64, (2001) 034007.}
 \bibitem{CB3}     {M.Loewe and C.Villavicencio, Phys.Rev. D67, (2003)074034; Phys.Rev.D70, (2004)074005}
 \bibitem{ladder}  {A.Barducci, R.Casalbuoni, G.Pettini, L.Ravagli, Phys.Lett.B564, (2003)217.}
 \bibitem{random}  {B.Klein, D.Toublan and J.J.M.Verbaarschot, Phys.Rev.D68, (2003)014009.}
 \bibitem{str}     {Yusuke Nishida, Phys.Rev.D69, (2004)094501}
 \bibitem{NJL1}    {D.Toublan and J.B.Kogut, Phys.Lett.B564, (2003)212.}
 \bibitem{NJL2}    {A.Barducci, R.Casalbuoni, G.Pettini, L.Ravagli, Phys.
                    Rev.D69, (2003)096004;  Phys.Rev.D71, (2005)016011}
 \bibitem{NJL3}    {M.Frank, M.Buballa and M.Oertel, Phys.Lett.B562, (2003)221.}
 \bibitem{NJL3-1}   {L.He and P.Zhuang, Phys.Lett.B615, (2005)93.}
 \bibitem{NJL3-2}   {L.He, M.Jin and P.Zhuang, Phys.Rev.D71, (2005)116001.}
 \bibitem{NJL4}     {Y.Nambu and G.Jona-Lasinio, Phys.Rev.
                    122, (1961)345;124, (1961)246.}
 \bibitem{NJL5}    {For reviews and general references, see U.Vogl and W.Weise, Prog.
                     Part. and Nucl.Phys.27, (1991)195; S.P.Klevansky, Rev.
                     Mod.Phys.64, (1992)649; M.K.Volkov, Phys.Part.Nucl.24, (1993)35; T.Hatsuda and T.Kunihiro, Phys.Rept.247, (1994)338;
                     M.Buballa, Phys.Rept.407, (2005)205}
 \bibitem{NJL6}    {J.Hufner, S.P.Klevansky, P.Zhuang and H.Voss, Ann.Phys(N.Y)
                    234, (1994)225; P.Zhuang, J.Hufner and S.P.Klevansky, Nucl.Phys.A576, (1994)525.}
 \bibitem{NJL7}    {P.Zhuang and Z.Yang, Phys.Rev.C62, (2000)054901.}
 \bibitem{LA2}     {X.Q.Luo, Phys.Rev.D70, (2004)091504(R); H.S.Chen and X.Q.Luo, Phys.Rev.D72, (2005)034504; E.B.Gregory, S.H.Guo, H.Kroger and X.Q.Luo, Phys.Rev.D62, (2000)054508.}
 \bibitem{NJL8}    {T.M.Schwarz, S.P.Klevansky, and G.Rapp, Phys.Rev.C60, (1999)055205.}
 \bibitem{NJL9}    {M.Huang, P.Zhuang, and W.Chao, Phys.Rev.D65, (2002)076012.}
 \bibitem{NJL10}   {M.Huang, P.Zhuang, and W.Chao, Phys.Rev.D67, (2003)065015.}
 \bibitem{NJL11}   {J.Liao, and P.Zhuang, Phys.Rev.D68, (2003)114016.}
 \bibitem{NJL12}   {I.Shovkovy, and M.Huang, Phys.Lett.B564, (2003)205.}
 \bibitem{Hooft}   {G.t'Hooft, Phys.Rev.D14, (1976)3432;Phys.Rept.142, (1986)357}


\end{thebibliography}
\end{document}